\documentclass[twocolumn,10pt,preprintnumbers,amsmath,amssymb]{revtex4}

\usepackage{epsf}
\usepackage{graphicx}  
\usepackage{dcolumn}   
\usepackage{bm}        

\newcommand{\be}{\begin{equation}}
\newcommand{\en}{\end{equation}}
\newcommand{\ba}{\begin{array}}
\newcommand{\ea}{\end{array}}
\newcommand{\bea}{\begin{eqnarray}}

\newcommand{\ena}{\end{eqnarray}}

\begin{document}

\preprint{BUN/1022-2006}

\title{Crossing $w=-1$ by a single scalar on a Dvali-Gabadadze-Porrati brane
 }

\author{Hongsheng Zhang and
  Zong-Hong Zhu\footnote{E-mail address: zhuzh@bnu.edu.cn}}
\affiliation{
 Department of Astronomy, Beijing Normal University, Beijing 100875, China}


\begin{abstract}
Recent type Ia supernovae data seem to favor a dark energy model
whose equation of state $w(z)$ crosses $-1$, which is a much more
amazing problem than the acceleration of the universe. Either the
case that $w(z)$ evolves from above $-1$ to below $-1$  or the case
that $w(z)$ runs from   below $-1$ to above $-1$,
 is consistent with
 present data. In this paper we show that it is possible to realize
 the crossing behaviours of both of the two cases  by only a single scalar field
 in frame of Dvali-Gabadadze-Porrati braneworld. At the same time we prove
 that there does not exist  scaling solution in a universe with
 dust.
\end{abstract}


 \maketitle

\section{Introduction}

 The existence of dark energy is one of the most significant cosmological discoveries
 over the last decades \cite{acce}. However, the nature of this dark energy remains a
 mystery. Various models of dark energy have been proposed, such as a small positive
 cosmological constant, quintessence, k-essence, phantom, holographic dark energy,
 etc., see \cite{review} for
 recent reviews with fairly complete list of references of different dark energy
 models. A cosmological constant is a  simple candidate for dark
  energy. However, following the more accurate data a more dramatic result
  appears:
  the recent analysis of the type Ia supernovas data
  indicates that the time varying dark energy gives a better
  fit  than a cosmological constant, and in particular, the equation of state (EOS)
   parameter
   $w$ (defined as the ratio of
 pressure to energy density) may cross $-1$ \cite{vari}.
 The dark energy with $w<-1$ is called phantom dark
 energy~\cite{call}, for which all energy conditions are
 violated. Here, it should be noted that the possibility that the dark energy behaves as
  phantom today is yet a matter under debate: the
 observational data, mainly those coming from the type Ia supernovae
 of high redshift and cosmic microwave background, may lead to different conclusions
 depending on what
 samples are selected, and what statistical analysis is
 applied \cite{jassal}. By contrast, other researches imply that all
 classes of dark energy models are comfortably allowed  by those
 very observations \cite{contra}. Presently all observations
 seem to  not rule out the possibility of the existence of matter with $w
 <-1$. Even in a model in which the Newton constant is evolving with
 respect redshift $z$, the best fit w(z) crosses the phantom divide $w=-1
 $ \cite{Nesseris}. Hence the phenomenological model for phantom dark energy
 should be considered seriously.
  To obtain $w <-1$, scalar field with a negative kinetic term,
  may be a simplest realization. The model with phantom matter has been investigated
  extensively \cite{phantom}, and a test of such matters in solar system, see \cite{solar}.  However,
 the EOS of phantom  field is  always
 less than $-1$  and can not cross $-1$.  It is easy to understand that if we put 2
 scalar fields into the model, one is an ordinary scalar and the other
 is a phantom: they dominate the universe by turns, under this
 situation the effective EOS can cross $-1$ \cite{quintom}. As we
 have mentioned that either the
 case that $w(z)$ evolves from above $-1$ to below $-1$ (Case I) or
 the case that $w(z)$ runs from   below $-1$ to above $-1$ (Case
 II) is consistent with
 present data.
   The observation data
  mildly favor dark energy of Case I , but also leave enough space for a
   dark energy of Case II \cite{III}.  It is worthy to point out that
  there  exists some interacting models, in which the effective EOS of dark energy
  crosses $-1$ \cite{interact}.
   More recently
  it has been found that crossing $-1$ within one scalar field
  model is possible, the cost is the action contains higher derivative
  terms~\cite{Li} (see also \cite{Binflation}). Also it is found
  that such a crossing can be realized
   without introducing ordinary scalar or phantom component in a
 Gauss-Bonnet brane world with induced gravity, where a four
 dimensional curvature scalar on the brane and a five dimensional
 Gauss-Bonnet term in the bulk are present \cite{hs1}. Other two
 related works on phantom divide crossing can be found in
 \cite{kamen}.

 In this paper we suggest a possibility with effective EOS
 crossing $-1$ in brane world scenario with only a single scalar field.
  The brane world
 scenario is now one of the most important ideas in high energy
 physics and cosmology. In this scenario, the standard model particles are
 confined to the 3-brane, while the gravitation can propagate in
 the whole spacetime. As for cosmology in the brane world scenario,
 many works have been done over the last several years; for a
 review, see \cite{maarten} and references therein. Brane world
 models admit a much wider range of possibilities for dark
 energy~\cite{sahni}. In an interesting braneworld model  proposed by Dvali,
  Gabadadze and Porrati (DGP)~\cite{dgpmodel} a
 late-time self-acceleration solution~\cite{dgpcosmology} appears naturally.
  In the DGP model, the bulk is a flat Minkowski spacetime, but a reduced
  gravity term appears on the brane, which is tensionless.
  In this model, gravity appears
  4-dimensional at short distances but is altered at distance larger
  than some freely adjustable crossover scale $r_c$ through the
 slow evaporation of the graviton off our 4-dimensional brane world
 universe into an unseen, yet large, fifth dimension. The late-time
 acceleration is driven by the manifestation of the excruciatingly
 slow leakage of gravity off our four-dimensional world into an extra
 dimension. In this scenario the universe eventually evolves into a de Sitter
 phase, and the effective EOS of dark energy never comes across
 $-1$. The
  behaviour of crossing $w=-1$ of the effective dark energy on a
  DGP brane with a cosmological constant and dust has been
  studied in \cite{ruth}.

 In the present paper we study some interesting properties of the
 scalar, including ordinary scalar and phantom, as dark energy, in the late time
 universe. We find effective EOS of the dark energy can cross $-1$
 only by a single scalar: For ordinary scalar, the EOS runs from above $-1$ to below $-1$
 ,evolving as dark energy of Case I in the negative branch,
  and for phantom, the EOS runs from below $-1$ to above $-1$, evolving as
  dark energy of Case II
  in the positive branch. As a byproduct, we prove that both for
  ordinary scalar and phantom, there does not exist
  scaling solution in a universe with dust in DGP brane world
  scenario.

  In the next section we shall investigate our model in detail. And
  in section III, we present the main conclusions and some
  discussions.

\section{The Model}
 Let us start from the action of the DGP model
 \be
 \label{totalaction}
 S=S_{\rm bulk}+S_{\rm brane},
 \en
where
 \be
 \label{bulkaction}
  S_{\rm bulk} =\int_{\cal M} d^5X \sqrt{-{}^{(5)}g}
  {1 \over 2 \kappa_5^2} {}^{(5)}R ,
 \en
and
 \be
 \label{braneaction}
 S_{\rm brane}=\int_{M} d^4 x\sqrt{-g} \left[
{1\over\kappa_5^2} K^\pm + L_{\rm brane}(g_{\alpha\beta},\psi)
\right].
 \en
Here $\kappa_5^2$ is the  5-dimensional gravitational constant,
${}^{(5)}R$ is the 5-dimensional curvature scalar and the matter
Lagrangian in the bulk. $x^\mu ~(\mu=0,1,2,3)$ are the induced
4-dimensional coordinates on the brane, $K^\pm$ is the trace of
extrinsic curvature on either side of the brane and $L_{\rm
brane}(g_{\alpha\beta},\psi)$ is the effective 4-dimensional
Lagrangian, which is given by a generic functional of the brane
metric $g_{\alpha\beta}$ and matter fields $\psi$ on the brane.

Consider the brane Lagrangian consisting of the following terms
\begin{eqnarray}
\label{lbrane}
 L_{\rm brane}=  {\mu^2 \over 2} R  + L_{\rm
m}+L_{\phi},
\end{eqnarray}
where $\mu$ is 4-dimensional reduced Planck mass, $R$ denotes the
curvature scalar on the brane, $L_{\rm m}$ stands for the Lagrangian
of other matters on the brane, and $L_{\phi}$ represents the
lagrangian of a scalar confined to the brane.
 Then assuming a  mirror symmetry in the bulk, we
  have the Friedmann equation on the brane \cite{dgpcosmology},
  see also \cite{hs1},
  \bea
 H^2+\frac{k}{a^2}=\frac{1}{3\mu^2}\left[\rho+\rho_0+\theta\rho_0
 (1+\frac{2\rho}{\rho_0})^{1/2}\right]
 \label{fried}
 \ena
 where $H=\dot{a}/a$ is the Hubble parameter, $a$ is the
 scale factor, $k$ is the spatial curvature of the
  three dimensional maximally symmetric space in the FRW metric on the brane,
  and $\theta=\pm 1$ denotes the two branches of DGP model, $\rho$ denotes
  the total energy density, including dust matter and scalar, on the brane,
  \be
  \rho=\rho_{\phi}+\rho_{dm},
  \en
  and the term $\rho_0$ relates the the strength of the 5-dimensional
  gravity with respect to 4-dimensional gravity,
 \be
 \rho_0=\frac{6\mu^2}{r_c^2},
 \en
 where the cross radius is defined as $r_c\triangleq \kappa_5^2\mu^2$.

  In quintom model \cite{quintom}, two scalar fields must be introduced  for the
  crossing $-1$ of effective EOS of dark energy. By contrast,
  we shall show that in our model only one field  is enough for this crossing
  behaviour by aiding of the 5-dimensional gravity.
  Therefore, the accelerated expansion
  of the universe is due to the combined effect of the scalar and
  the competition between 4-dimensional gravity and the 5-dimensional gravity.

 To explain the the observed evolving of EOS of effective dark energy,
  we calculate the equation of state $w$ of the effective
  ``dark energy" caused by the scalar field and term representing
  brane world effect
  by comparing the modified Friedmann equation in
  the brane world scenario and the standard Friedmann equation in general
  relativity, because all observed features of dark energy are
  ``derived" in general relativity.
   Note that the Friedmann equation in the
 four dimensional
  general relativity can be written as
 \be
 H^2+\frac{k}{a^2}=\frac{1}{3\mu^2} (\rho_{dm}+\rho_{de}),
 \label{genericF}
 \en
 where the first term of RHS of the above equation represents the dust matter and the second
 term stands for the effective dark energy. Compare (\ref{genericF})
 with (\ref{fried}), one obtains the density of effective dark
 energy,
 \be
 \rho_{de}=\rho_{\phi}+\rho_0+\theta\rho_0\left[\rho+\rho_0+\theta\rho_0
 (1+\frac{2\rho}{\rho_0})^{1/2}\right].
 \label{rhode}
 \en

 Since the dust matter obeys the continuity equation
 and the Bianchi identity keeps valid, dark energy itself satisfies
  the continuity equation
 \be
 \frac{d\rho_{de}}{dt}+3H(\rho_{de}+p_{eff})=0,
 \label{em}
 \en
 where $p_{eff}$ denotes the effective pressure of the dark energy.
 And then we can express the equation of state for the dark
 energy as
   \be
  w_{de}=\frac{p_{eff}}{\rho_{de}}=-1+\frac{1}{3}\frac{d \ln \rho_{de}}{d \ln
  (1+z)},
   \en
   where, from (\ref{em}),
   \bea
   \nonumber
   \frac{d \ln \rho_{de}}{d \ln
  (1+z)}= \frac{3}{\rho_{de}}\left[\rho_{de}+p_{de}~~~~~~~~~~~~~~~~~~~~~~~~~ \right.\\
  \left.~~~~~~~~+\theta(1+2\frac{\rho_{de}+\rho_{dm}}{\rho_0})
  ^{-1/2}(\rho_{de}+\rho_{dm}+p_{de})\right],
  \label{derivation}
  \ena
  where $p_{de}$ represents the pressure of the dark energy. Here we
  stress that it is different from $p_{eff}$.
  Clearly, if $\frac{d \ln \rho_{de}}{d \ln
  (1+z)}$ is greater than 0, dark energy evolves as phantom; if $\frac{d \ln \rho_{de}}{d \ln
  (1+z)}$ is less than 0, it evolves as quintessence; if $\frac{d \ln \rho_{de}}{d \ln
  (1+z)}$ equals 0, it is just cosmological constant. In a more
  intuitionistic way, if $\rho_{de}$ decreases and then increases
  with respect to redshift (or time), or increases and then
  decreases, which implies that EOS of dark energy crosses phantom divide.
 It is known that the equations of sort (\ref{fried}) and (\ref{em}) belong
 to so-called inhomogeneous EOS of FRW universe \cite{inho} or FRW universe
 with general EOS \cite{frwg}, which can cross the phantom divide.

   In the DGP brane world model, the induced gravity correction arises because the
 localized massive scalars fields on the brane, which couple to bulk
 gravitons, can generate via quantum loops a localized
 four-dimensional world-volume kinetic term for gravitons~\cite{origin}.
 Therefore, to investigate the behaviour of classical level of such a
 field is not only interesting for cosmology, but also important for
 analysesing the DPG model itself. The behaviour of the scalar in the early universe
 ,ie, the inflation driven by a scalar
 field confined to a DGP brane, has been investigated in \cite{hs2}
 \cite{pz}. A proposal that the
   same phantom scalar plays the role of early
 time (phantom) inflaton and late time dark energy is presented in
 \cite{inflaton}. In view of these progresses, it is valuable to
 study the behaviour of a scalar field on a DGP brane at late time
 universe.

   In the following two subsections, we shall analyze the dynamics
   of an ordinary scalar and a phantom in the late time universe on
   a  DGP brane, respectively. We show both of them can cross the
   phantom divide: in the negative branch for an ordinary scalar; in the
   positive branch for a phantom.
   \subsection{Ordinary scalar field}
  For an ordinary scalar, the action in $L_{\phi}$ of (\ref{lbrane})
  reads,
  \be
  L_{\phi}=-\frac{1}{2}\partial_\mu \phi
  \partial^\mu \phi-V(\phi).
  \en
  Varying the action with respect to the metric tensor in an FRW universe we reach to
  \bea
  \rho_{\phi}=\frac{1}{2}\dot{\phi}^2+V(\phi),\\
  p_{\phi}=\frac{1}{2}\dot{\phi}^2-V(\phi),
  \ena
  where a dot denotes derivative with respect to time.
  The exponential potential is an important
 example which can be solved exactly in the standard model. Also it
 has been shown that the inflation driven by a scalar with
  exponential potential can exit naturally in the warped DGP model \cite{hs1}.
    In
 addition, we know that such exponential potentials of scalar fields
 occur naturally in some fundamental theories such as string/M
 theories.  It is therefore quite interesting to
 investigate a scalar with such a potential in late time universe on a DGP brane.
 Here we set
 \be
 V=V_0e^{-\lambda\frac{\phi}{\mu}}.
 \en
 Here $\lambda$ is a constant and $V_0$ denotes the initial value of
 the potential.

 Then the derivation of effective density of dark energy with respective to $\ln (1+z)$
  reads,
 \bea
 \nonumber
 \frac{d \rho_{de}}{d \ln
  (1+z)}&=&3[\dot{\phi}^2\\
  +\theta(1&+&\frac{\dot{\phi}^2+2V+2\rho_{dm}}{\rho_0})
  ^{-1/2}(\dot{\phi}^2+\rho_{dm})].
  \label{derivation or}
  \ena
  If $\theta=1$, both terms of RHS are positive, hence it never goes
  to zero at finite time. But if $\theta=-1$, the two terms of RHS
  carry opposite sign, therefore it is possible that the EOS of dark
  energy crosses phantom divide. In the following of this subsection
  we only consider the case of $\theta=-1$.
  For a more detailed research of the evolution of the variables in this model
  we write them in a dynamical system, which can be derived from the Friedmann equation
  (\ref{fried}) and continuity equation (\ref{em}). We first define some new
  dimensionless variables,
  \bea
  x&\triangleq&\frac{\dot{\phi}}{\sqrt{6}\mu H},\\
  y&\triangleq&\frac{\sqrt{V}}{\sqrt{3}\mu H},\\
  l&\triangleq&\frac{\sqrt{\rho_m}}{\sqrt{3}\mu H},\\
  b&\triangleq&\frac{\sqrt{\rho_0}}{\sqrt{3}\mu H}.
  \ena
  The dynamics of the universe can be described by the following
  dynamical system with these new dimensionless variables,
 \bea
 \label{1}
 x'&=&-\frac{3}{2}\alpha
 x(2x^2+l^2)+3x-\frac{\sqrt{6}}{2}\lambda y^2,\\
 \label{2}
  y'&=&-\frac{3}{2}\alpha
 y(2x^2+l^2)+\frac{\sqrt{6}}{2}\lambda xy,\\
 \label{3}
   l'&=&-\frac{3}{2}\alpha
 l(2x^2+l^2)+\frac{3}{2}l,\\
 \label{4}
  b'&=&-\frac{3}{2}\alpha
 b(2x^2+l^2),
 \ena
 where
 \be
 \alpha\triangleq 1-\left(1+2\frac{x^2+y^2+l^2}{b^2}\right)^{-1/2},
 \en
 a prime stands for derivation with respect to
 $s\triangleq-\ln(1+z)$, and we have set $k=0$, which is implied
 either by theoretical side (inflation in the early universe)
 ,or observation side (CMB fluctuations \cite{WMAP}). One can check this system degenerates to a
 quintessence with dust matter in standard general relativity.
  Note that the 4 equations (\ref{1}), (\ref{2}),
 (\ref{3}), (\ref{4}) of this system are not independent. By using the Friedmann
 constraint, which can be derived from the Friedmann equation,
 \be
 \label{constraint}
 x^2+y^2+l^2+b^2-b^2\left(1+2\frac{x^2+y^2+l^2}{b^2}\right)^{1/2}=1,
 \en
 the number of the independent equations can be reduced to 3.
    There
 are
 two critical points of this system satisfying $x'=y'=l'=b'=0$ appearing at
 \bea
 &x&=y=l=0, ~~b={\rm constant};\\
 &x&=y=l=b=0.
 \ena
 However, neither of them satisfies the Friedmann constraint
 (\ref{constraint}). Hence we prove that there is no
 kinetic energy-potential energy scaling solution or kinetic energy-potential energy
 -dust matter scaling solution on a DGP brane with quintessence and dust.
 Since the equation set (\ref{1}), (\ref{2}), (\ref{3}), (\ref{4})
 can not be solved analytically, we present
 some numerical results about the dark energy density. And one will see that in
 reasonable regions of parameters, the EOS of dark energy crosses
 $-1$.

\begin{figure}
\centering
 \includegraphics[totalheight=1.8in, angle=0]{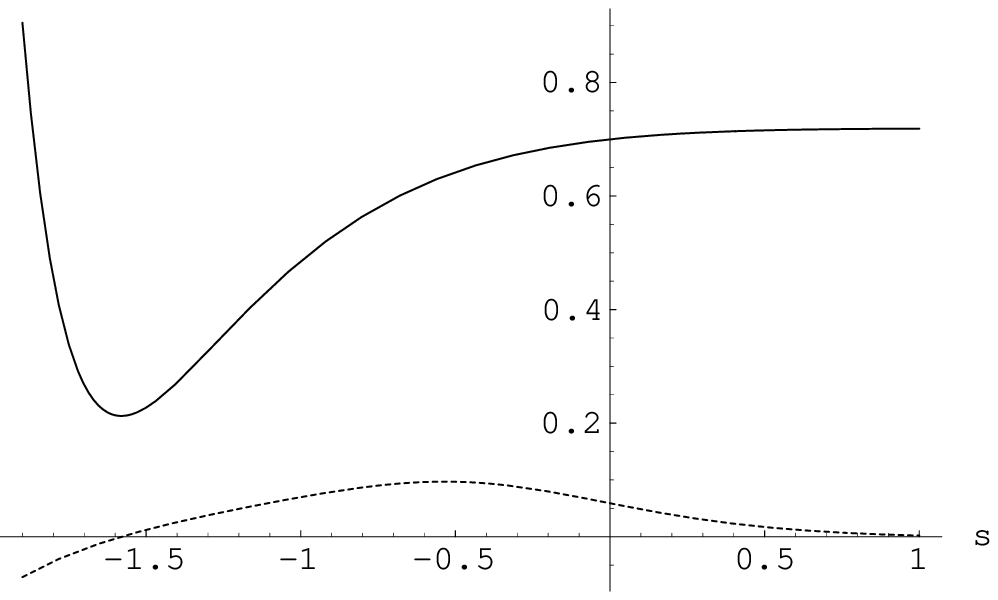}
\includegraphics[totalheight=2in, angle=0]{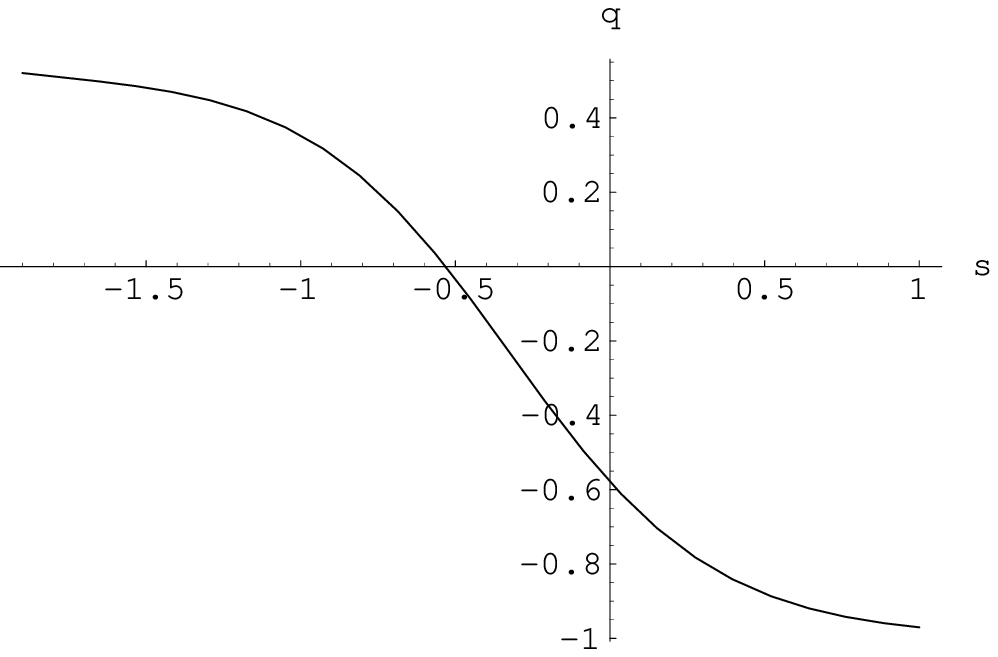}
\caption{For this figure, $\Omega_{ki}=0.01$, $\Omega_{r_c}=0.01$,
$\lambda=0.05$. {\bf{(a)}} $\beta$ and $\gamma$ as functions of $s$,
in which $\beta$ resides on the solid line, while $\gamma$ dwells at
the dotted line. The EOS of dark energy crosses $-1$ at about
$s=-1.6$, or $z=3.9$. {\bf{(b)}} The corresponding deceleration
parameter, which crosses 0 at about $s=-0.52$, or $z=0.68$.}
 \label{rho1}
 \end{figure}

\begin{figure}
\centering
 \includegraphics[totalheight=1.8in, angle=0]{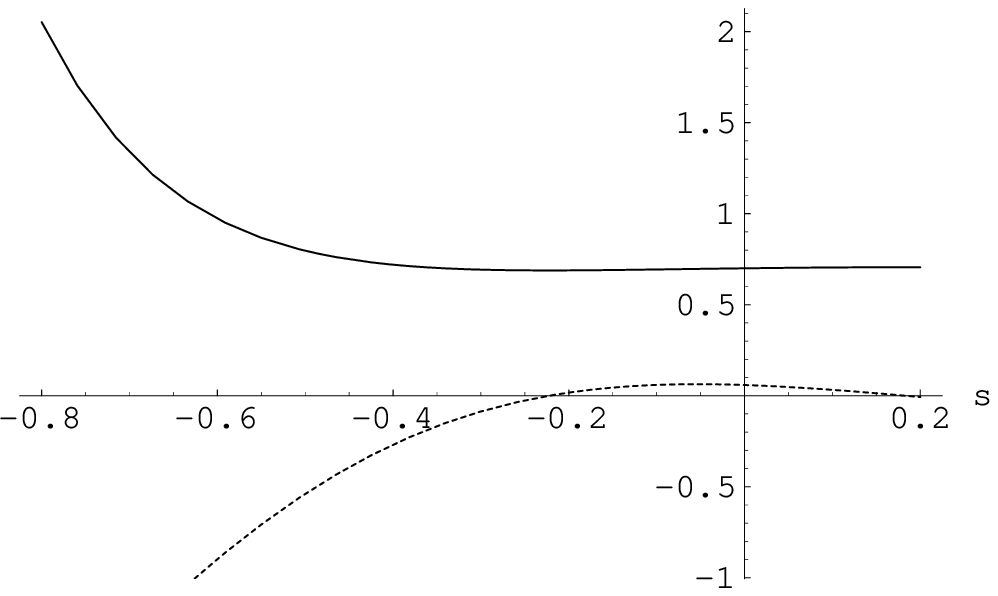}
\includegraphics[totalheight=2in, angle=0]{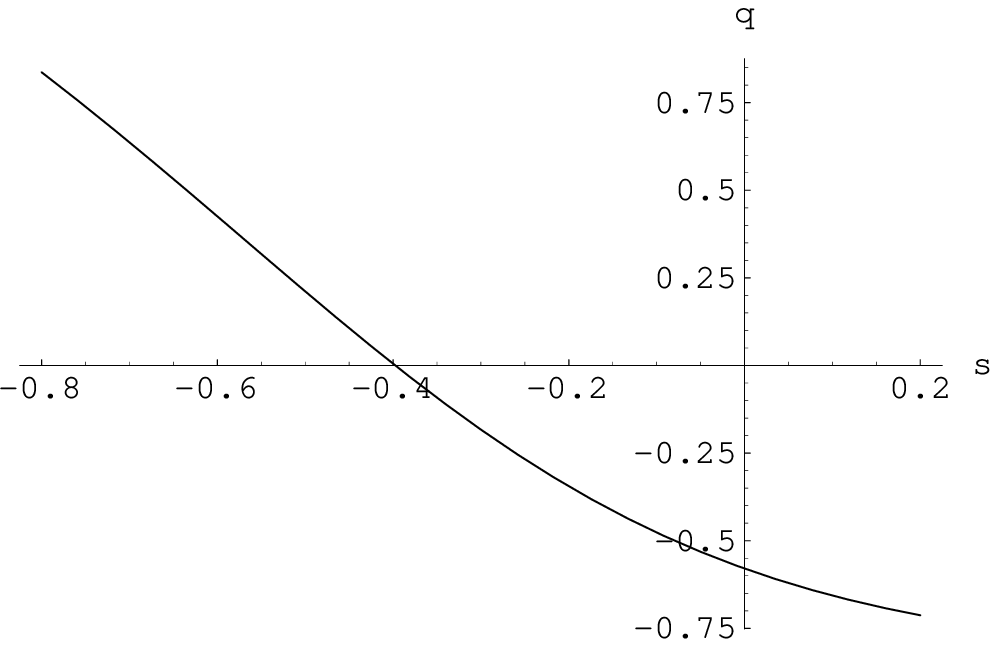}
\caption{For this figure, $\Omega_{ki}=0.01$, $\Omega_{r_c}=0.01$,
$\lambda=0.5$. {\bf{(a)}} $\beta$ and $\gamma$ as functions of $s$,
in which $\beta$ resides on the solid line, while $\gamma$ dwells at
the dotted line. The EOS of dark energy crosses $-1$ at about
$s=-0.22$, or $z=0.25$. {\bf{(b)}} The corresponding deceleration
parameter, which crosses 0 at about $s=-0.40$, or $z=0.49$.}
 \label{rho2}
 \end{figure}

\begin{figure}
\centering
 \includegraphics[totalheight=1.8in, angle=0]{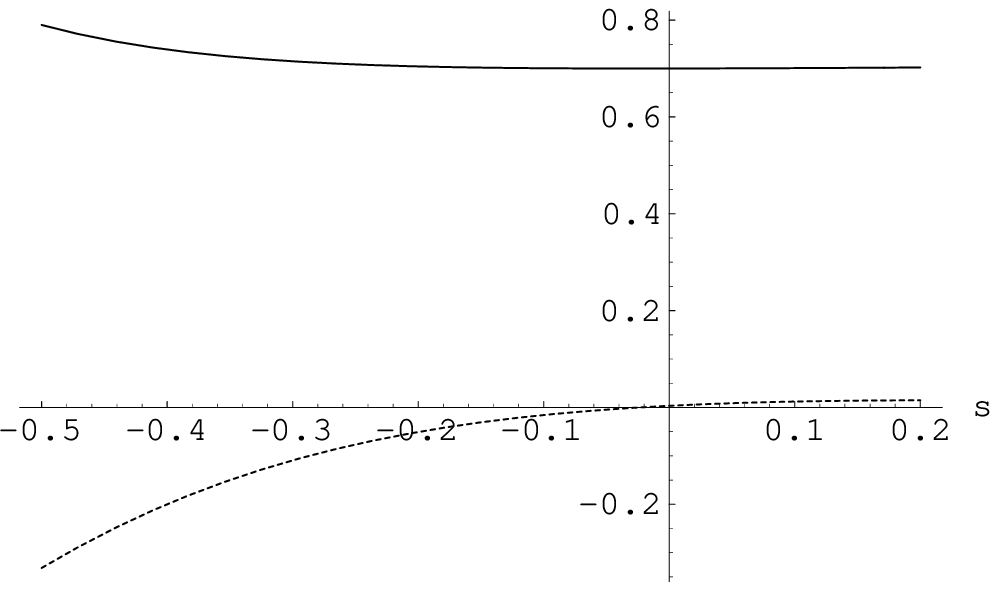}
\includegraphics[totalheight=2in, angle=0]{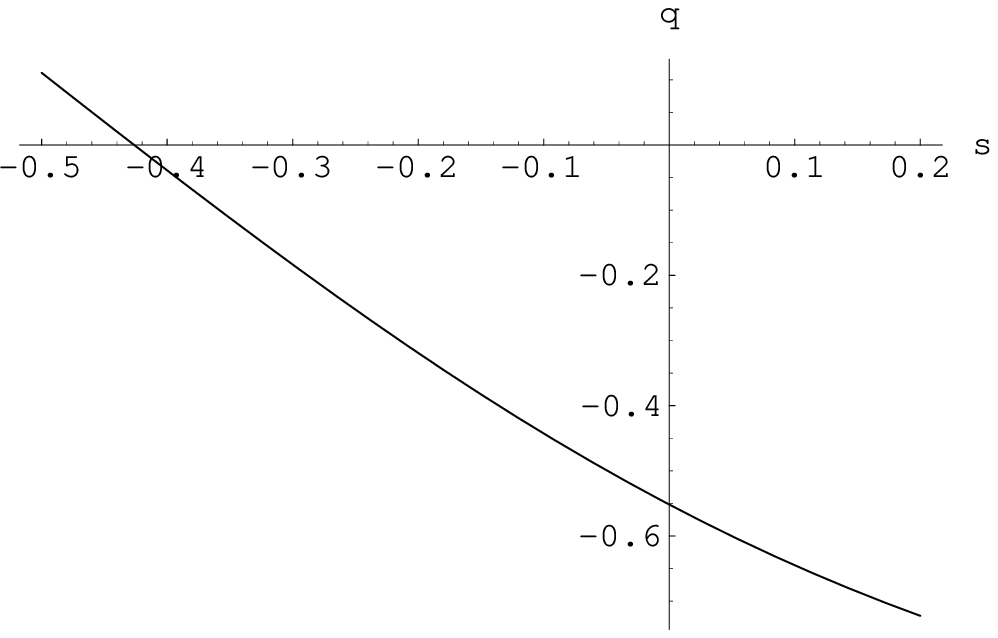}
\caption{For this figure, $\Omega_{ki}=0.1$, $\Omega_{r_c}=0.01$,
$\lambda=0.05$. {\bf{(a)}} $\beta$ and $\gamma$ as functions of $s$,
in which $\beta$ resides on the solid line, while $\gamma$ dwells at
the dotted line. The EOS of dark energy crosses $-1$ at about
$s=-0.02$, or $z=0.02$. {\bf{(b)}} The corresponding deceleration
parameter, which crosses 0 at about $s=-0.41$, or $z=0.51$.}
 \label{rho3}
 \end{figure}

 The stagnation point of $\rho_{de}$ dwells at
 \be
 \frac{b}{\sqrt{2}+b}\left(2+\frac{l^2}{x^2}\right)=2,
 \en
 which can be derived from (\ref{derivation or}) and
 (\ref{constraint}). One concludes from the above equation that a smaller $r_c$, a
 smaller $\Omega_m$ (which is defined as the present value of
  the energy density of dust matter over the critical density), or
  a larger $\Omega_{ki}$ (which is defined as the present value of
  the kinetic energy density of the scalar  over the critical
  density) is helpful to shift the stagnation point to lower redshift
  region. We show a concrete numerical example of this crossing
  behaviours in Fig. \ref{rho1}, Fig. \ref{rho2}, Fig. \ref{rho3}.
  For convenience we introduce the dimensionless density and rate of
  change with respect to redshift of dark energy as below.
 \bea
 \nonumber
 \beta=\frac{\rho_{de}}{\rho_c}=\left.\frac{\Omega_{r_c}}{b^2}\right[x^2+y^2+b^2
  \\
 \left. -b^2(1+2\frac{x^2+y^2+l^2}{b^2})^{1/2}\right],
 \ena
 where $\rho_c$ denotes the present critical density of the universe, and
 \bea
 \nonumber
 \gamma &=&
 \frac{1}{\rho_c}\frac{b^2}{\Omega_{r_c}}\frac{d\rho_{de}}{ds}\\
 &=&3\left[(1+2\frac{x^2+y^2+l^2}{b^2})^{-1/2}(2x^2+l^2)-2x^2\right].
 \ena
  The most significant parameters from the viewpoint of
  observations is the deceleration parameter $q$, which carries the total
  effects of cosmic fluids. $q$ is defined as
  \bea
  \nonumber
  q&=&-\frac{\ddot{a}a}{\dot{a}^2}\\
   &=&-1+\frac{3}{2}\alpha(2x^2+l^2),
   \ena
   which is also plotted in these figures for corresponding density
   curve of dark energy. In all the figures we set $\Omega_m=0.3$
   but with different $\Omega_{ki}$, $\lambda$, $\Omega_{r_c}$, in
   which $\Omega_{r_c}$ is defined as
   the present value of
  the energy density of $\rho_0$ over the critical density
  $\Omega_{r_c}={\rho_0}/{\rho_c}$.

 From Fig 1, 2, and 3, clearly, the EOS of effective dark energy
 crosses $-1$ as expected. At the same time the deceleration
 parameter is consistent with observations. As is well known,
 the EOS of a single scalar in standard general relativity never
 crosses the phantom divide, the induced term, through the ``energy density" of $r_c$,
 $\rho_0$,
 plays a critical in this crossing. We see that a small component of
 $\rho_0$, i.e. $\Omega_{r_c}=0.01$, is successfully to make the EOS
 of dark energy cross $-1$. The other note is that here the EOS runs from
 above $-1$ to below $-1$
 ,which describes the dark energy of Case I.

\subsection{Phantom field}
 In this subsection we investigate a phantom field in DGP model.
 Most of the discussions are parallel to the last subsection.
  For a phantom field, the action in $L_{\phi}$ of (\ref{lbrane})
  reads,
  \be
  L_{\phi}=\frac{1}{2}\partial_\mu \phi
  \partial^\mu \phi-V(\phi).
  \en
  A variation of the action with respect to the metric tensor in an FRW universe
  yields,
  \bea
  \rho_{\phi}=-\frac{1}{2}\dot{\phi}^2+V(\phi),\\
  p_{\phi}=-\frac{1}{2}\dot{\phi}^2-V(\phi).
  \ena
  To compare with the results of the ordinary scalar, here we set a
  same potential as before,
 \be
 V=V_0e^{-\lambda\frac{\phi}{\mu}}.
 \en

 The ratio of change of effective density of dark energy with respective to $\ln (1+z)$
  becomes,
 \bea
 \nonumber
 \frac{d \rho_{de}}{d \ln
  (1+z)}&=& 3[-\dot{\phi}^2+\theta(1\\
  +\frac{-\dot{\phi}^2+2V+2\rho_{dm}}{\rho_0})
  &^{-1/2}&(-\dot{\phi}^2+\rho_{dm})].
  \label{derivation ph}
  \ena
  To study the behaviour of the EOS of dark energy, we first take a
  look at the signs of the terms of RHS of the above equation.
  $(-\dot{\phi}^2+\rho_{dm})$ represents the total energy density of
  the cosmic fluids, which should be positive. The term $(1
  +\frac{-\dot{\phi}^2+2V+2\rho_{dm}}{\rho_0})
  ^{-1/2}$ should also be positive. Hence if $\theta=-1$,
   both terms of RHS are negative: It never goes
  to zero at finite time. Contrarily, if $\theta=1$, the two terms of RHS
  carry opposite sign: The EOS of dark
  energy is able to cross phantom divide. In the following of the present subsection
  we consider the branch of $\theta=1$.

  Similar to the case of an ordinary scalar, the dynamics of the
  universe can be described by the following
  dynamical system,
 \bea
 \label{5}
 x'&=&-\frac{3}{2}\alpha'
 x(-2x^2+l^2)+3x-\frac{\sqrt{6}}{2}\lambda y^2,\\
 \label{6}
  y'&=&-\frac{3}{2}\alpha'
 y(-2x^2+l^2)+\frac{\sqrt{6}}{2}\lambda xy,\\
 \label{7}
   l'&=&-\frac{3}{2}\alpha'
 l(-2x^2+l^2)+\frac{3}{2}l,\\
 \label{8}
  b'&=&-\frac{3}{2}\alpha'
 b(-2x^2+l^2),
 \ena
 where
 \be
 \alpha'\triangleq 1+\left(1+2\frac{-x^2+y^2+l^2}{b^2}\right)^{-1/2},
 \en
 , and we have also adopted the spatial flatness condition. The
 definitions of $x,~y,~l,~ b$ are the same as the last subsection.
 One can check this system degenerates to a
 phantom with dust matter in standard general relativity.
  Also the 4 equations (\ref{5}), (\ref{6}),
 (\ref{7}), (\ref{8}) of this system are not independent. Now the Friedmann
 constraint becomes
 \be
 \label{constraint ph}
 -x^2+y^2+l^2+b^2+b^2\left(1+2\frac{-x^2+y^2+l^2}{b^2}\right)^{1/2}=1,
 \en
 with which there are 3 independent equations left in this system.
   Through a similar analysis as the case of an ordinary scalar, we can prove
    that there is no
 kinetic energy-potential energy scaling solution or kinetic energy-potential energy
 -dust matter scaling solution on a DGP brane with phantom and dust.
 The equation set (\ref{5}), (\ref{6}), (\ref{7}), (\ref{8})
 can not be solved analytically, therefore here we give
 some numerical results. Again, one will see that in
 reasonable regions of parameters, the EOS of dark energy crosses
 $-1$, but from below $-1$ to above $-1$.

\begin{figure}
\centering
 \includegraphics[totalheight=1.8in, angle=0]{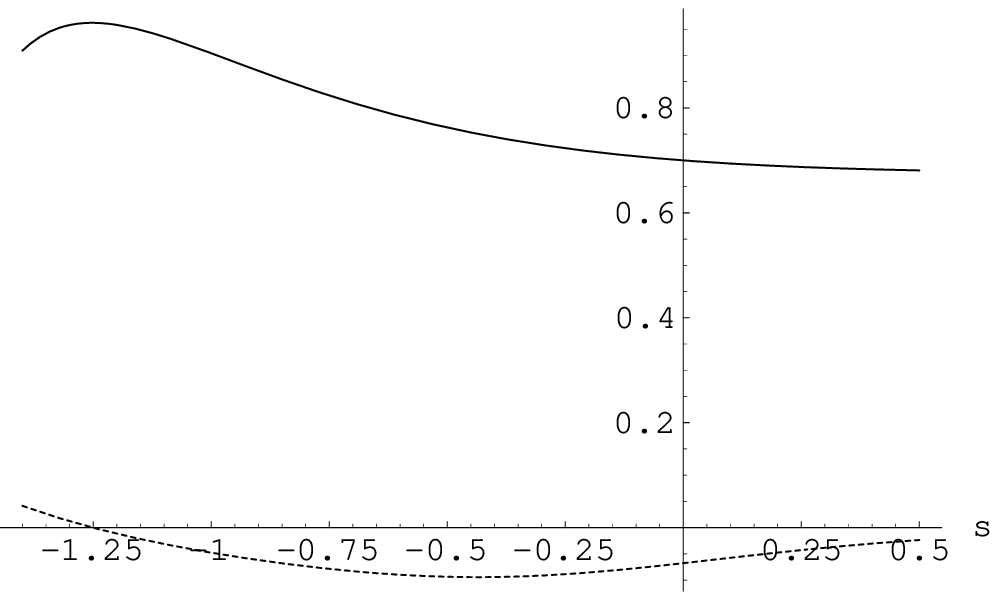}
\includegraphics[totalheight=2in, angle=0]{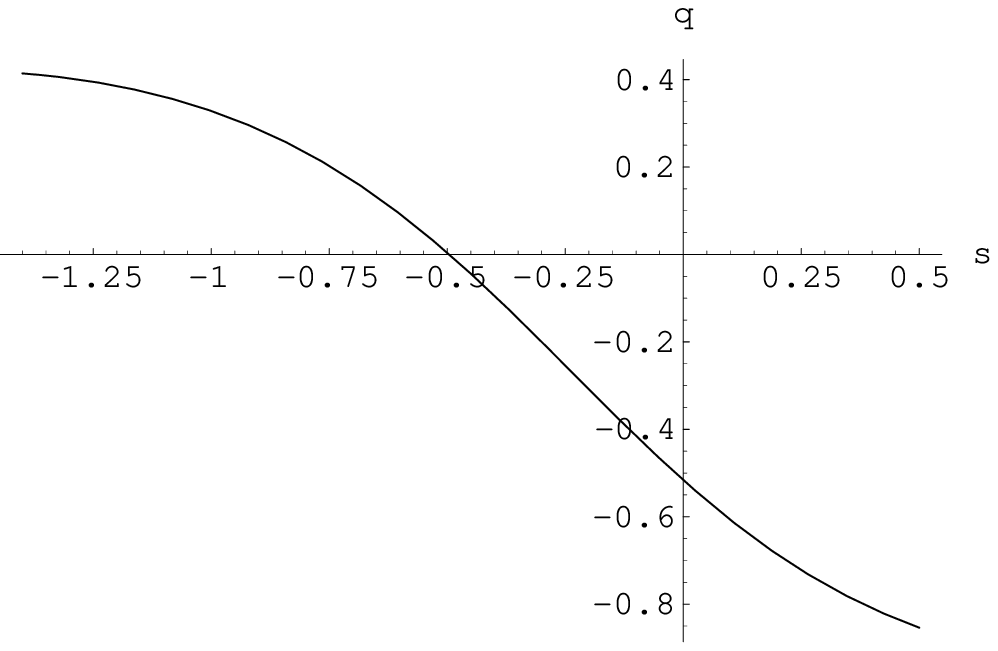}
\caption{For this figure, $\Omega_{ki}=0.01$, $\Omega_{r_c}=0.01$,
$\lambda=0.01$. {\bf{(a)}} $\beta$ and $\gamma$ as functions of $s$,
in which $\beta$ resides on the solid line, while $\gamma$ dwells at
the dotted line. The EOS of dark energy crosses $-1$ at about
$s=-1.25$, or $z=1.49$. {\bf{(b)}} The corresponding deceleration
parameter, which crosses 0 at about $s=-0.50$, or $z=0.65$.}
 \label{rho4}
 \end{figure}

 The stagnation point of $\rho_{de}$ inhabits at
 \be
 \frac{b}{\sqrt{2}-b}\left(-2+\frac{l^2}{x^2}\right)=2,
 \en
 which can be derived from (\ref{derivation ph}) and
 (\ref{constraint ph}). One concludes from the above equation that a smaller $r_c$, a
 smaller $\Omega_m$, or
  a larger $\Omega_{ki}$ is helpful to shift the stagnation point to lower redshift
  region, which is the same as the case of an ordinary scalar.
   Then we show a concrete numerical example of the crossing
  behaviour of this case in Fig. \ref{rho4}.
  The dimensionless density and rate of
  change with respect to redshift of dark energy become,
 \bea
 \nonumber
 \beta=\left.\frac{\Omega_{r_c}}{b^2}\right[-x^2+y^2+b^2
  \\
 \left. +b^2(1+2\frac{-x^2+y^2+l^2}{b^2})^{1/2}\right],
 \ena
 and
 \be
 \gamma =3\left[-(1+2\frac{-x^2+y^2+l^2}{b^2})^{-1/2}(-2x^2+l^2)+2x^2\right].
 \en
  The deceleration parameter $q$ becomes,
  \be
   q=-1+\frac{3}{2}\alpha'(-2x^2+l^2),
   \en
   which is plotted in the figure for corresponding density
   curve of dark energy. In this figures we also set $\Omega_m=0.3$.

 Fig. 4 explicitly illuminates that the EOS of effective dark energy
 crosses $-1$, as expected. At the same time the deceleration
 parameter is consistent with observations.
 The EOS of a single phantom field in standard general relativity always less
 than $-1$, therefore, the 5 dimensional gravity
 plays a critical role in this crossing, though
  we see that the ``geometric energy density" only takes a small
  component of the total density, i.e. $\Omega_{r_c}=0.01$. Finally, the most important
 difference of the dynamics between an ordinary scalar and the
 phantom of the crossing behaviour lies on the crossing manner:
 Here the EOS runs from
 below $-1$ to above $-1$
 , which can describe the dark energy of Case II, while the effective
  EOS of an ordinary scalar evolves from above $-1$ to below $-1$,
  as shown in subsection A.

\section{Conclusions and discussions}

 To summarize, this paper displays that in frame of DGP brane world
 it is possible to realize crossing  $w=-1$ for
 the EOS of the effective dark energy of a single scalar. For an
 ordinary scalar in the negative branch, the EOS transits from $w>-1$ to $w<-1$
 ,which can describe the dark energy of Case I, while for a phantom in the positive branch, the EOS
 transits from $w<-1$ to $w>-1$, which describes dark energy of Case II. The deceleration
 parameter can be consistent with observations.
 In standard general relativity, the EOS of a single scalar never
 crosses $-1$, therefore, the five dimensional gravity plays an
 important role in this transition, although the competent of this ``geometric energy
 density" $\rho_0$ over the critical energy density is very small.
 The other conclusion is that there does not exist scaling solution
 in a universe with dust on a DGP brane
  , neither in the positive branch nor in the negative branch.

  Fig. 2 shows that the deceleration parameter can exceed $0.5$ in some high
  redshift region, which implies that the scalar may enter a
  kinetic energy dominated phase. And hence the energy density of
  scalar will exceed the density of dust in such region, which can
  spoil the successful predictions of structure formation theory. Therefore,
  we should treat the exponential potential as an approximation of a
  more appropriate potential, such as the tracker potential \cite{tracker},
  in low redshift region. The present model may fail at very high
  redshift region (such as $z=1000$). How to construct a potential
  which can describe both the early universe and the late universe
  is  our further work.


{\bf Acknowledgments:}
This work was supported by
  the National Natural Science Foundation of China
    , under Grant No. 10533010, the Project-sponsored SRF for ROCS, SEM of China,
    and  Program for New Century Excellent Talents in University (NCET).


\begin{thebibliography}{99}



\bibitem{acce}
A. G. Riess et al. ,
Astron. J. 116, 1009 (1998), astro-ph/9805201; S. Perlmutter et
al.,
Astrophys. J. 517, 565 (1999), astro-ph/9812133.


\bibitem{review}
 Edmund J. Copeland, M. Sami and Shinji Tsujikawa,
 hep-th/0603057;
 J.~P.~Uzan,
  arXiv:astro-ph/0605313.

\bibitem{vari}
U.~Alam, V.~Sahni, T.~D.~Saini and A.~A.~Starobinsky,
  Mon.\ Not.\ Roy.\ Astron.\ Soc.\  {\bf 354}, 275 (2004)
  [arXiv:astro-ph/0311364];
  U.~Alam, V.~Sahni and A.~A.~Starobinsky,
  JCAP {\bf 0406}, 008 (2004)
  [arXiv:astro-ph/0403687];
 D. Huterer and A. Cooray, astro-ph/0404062;
 Y. Wang and M. Tegmark,
 astro-ph/0501351;
  Andrew R Liddle, Pia Mukherjee, David Parkinson and Yun Wang,
astro-ph/0610126.
 \bibitem{call}
 R.R. Caldwell,
 Phys.Lett. B545 (2002) 23, astro-ph/9908168;
 P.~Singh, M.~Sami and N.~Dadhich,
  Phys.\ Rev.\ D {\bf 68}, 023522 (2003)
  [arXiv:hep-th/0305110].


  \bibitem{jassal}
 H.~K.~Jassal, J.~S.~Bagla and T.~Padmanabhan,
  arXiv:astro-ph/0601389;
  H.~K.~Jassal, J.~S.~Bagla and T.~Padmanabhan,
  Phys.\ Rev.\ D {\bf 72} (2005) 103503
  [arXiv:astro-ph/0506748];
  S.~Nesseris and L.~Perivolaropoulos,
  arXiv:astro-ph/0610092.
  \bibitem{contra}
  V. Barger, E. Guarnaccia and D. Marfatia, Phys. Lett.
  B 635 (2006) 61 [arXiv:hep-ph/0512320].
  \bibitem{Nesseris}
  S.~Nesseris and L.~Perivolaropoulos,
  Phys.\ Rev.\ D {\bf 73} (2006) 103511
  [arXiv:astro-ph/0602053].

\bibitem{phantom}


   E.Elizalde, S.
Nojiri and S.D. Odintsov, Phys.\ Rev.\ D {\bf 70},2004,043539; S.
Nojiri and S. D. Odintsov, Phys.\ Lett.\ B {\bf 562},2003,147;B.
Boisseau, G. Esposito-Farese, D. Polarski, Alexei A. Starobinsky,
Phys. Rev. Lett. 85, 2236 (2000);R. Gannouji, D. Polarski, A.
Ranquet, A. A. Starobinsky JCAP 0609,016 (2006);

 Z.K. Guo, R.G. Cai and Y.Z. Zhang,
astro-ph/0412624; R.~G.~Cai and A.~Wang,
  JCAP {\bf 0503}, 002 (2005)
  [arXiv:hep-th/0411025];
  Z.K. Guo and Y.Z. Zhang,
   Phys.Rev. D71 (2005) 023501, astro-ph/0411524;
  Z.K. Guo, Y.S. Piao, X. Zhang and Y.Z. Zhang,
 Phys.Lett. B608 (2005) 177-182, astro-ph/0410654;
 S. M. Carroll, A. De Felice and M. Trodden
   Phys.Rev. D71 (2005)
 023525, astro-ph/0408081;
S. Nesseris and L. Perivolaropoulos
 Phys.Rev. D70 (2004)
 123529, astro-ph/0410309;
   Y.H. Wei
 , gr-qc/0502077;
S.~Nojiri, S.~D.~Odintsov and S.~Tsujikawa,
  Phys.\ Rev.\ D {\bf 71}, 063004 (2005)
  [arXiv:hep-th/0501025];
 P.~Singh,
  arXiv:gr-qc/0502086;
 H.~Stefancic,
  arXiv:astro-ph/0504518;
  V. K. Onemli and R. P. Woodard
  , Phys. Rev. D70:107301,2004;
  Class. Quant. Grav. 19 (2002) 4607;
  gr-qc/0612026.

 \bibitem{solar}
 J.~Martin, C.~Schimd and J.~P.~Uzan,
  Phys.\ Rev.\ Lett.\  {\bf 96}, 061303 (2006)
  [arXiv:astro-ph/0510208];







   \bibitem{quintom}
   Z.~K.~Guo, Y.~S.~Piao, X.~Zhang and Y.~Z.~Zhang,
  arXiv:astro-ph/0608165;
 B.~Feng, X.~L.~Wang and X.~M.~Zhang,
  Phys.\ Lett.\ B {\bf 607}, 35 (2005)
  [arXiv:astro-ph/0404224];
 M.~R.~Setare,
  Phys.\ Lett.\ B {\bf 641}, 130 (2006);
  X.~Zhang,
  Phys.\ Rev.\ D {\bf 74}, 103505 (2006)
  [arXiv:astro-ph/0609699];
  Y.~f.~Cai, H.~Li, Y.~S.~Piao and X.~m.~Zhang,
  arXiv:gr-qc/0609039;
  M.~Alimohammadi and H.~M.~Sadjadi,
  arXiv:gr-qc/0608016;
  H.~Mohseni Sadjadi and M.~Alimohammadi,
  Phys.\ Rev.\ D {\bf 74}, 043506 (2006)
  [arXiv:gr-qc/0605143];
  W.~Wang, Y.~X.~Gui and Y.~Shao,
  Chin.\ Phys.\ Lett.\  {\bf 23} (2006) 762;
  X.~F.~Zhang and T.~Qiu,
  arXiv:astro-ph/0603824;
  R.~Lazkoz and G.~Leon,
  Phys.\ Lett.\ B {\bf 638}, 303 (2006)
  [arXiv:astro-ph/0602590];
  X.~Zhang,
  Commun.\ Theor.\ Phys.\  {\bf 44}, 762 (2005);
  P.~x.~Wu and H.~w.~Yu,
  Int.\ J.\ Mod.\ Phys.\ D {\bf 14}, 1873 (2005)
  [arXiv:gr-qc/0509036];
  G.~B.~Zhao, J.~Q.~Xia, M.~Li, B.~Feng and X.~Zhang,
  Phys.\ Rev.\ D {\bf 72}, 123515 (2005)
  [arXiv:astro-ph/0507482];
  H.~Wei, R.~G.~Cai and D.~F.~Zeng,
  Class.\ Quant.\ Grav.\  {\bf 22}, 3189 (2005)
  [arXiv:hep-th/0501160];
  H.~Wei and R.~G.~Cai,
  Phys.\ Rev.\ D {\bf 72} (2005) 123507
  [arXiv:astro-ph/0509328];
  H.~Wei and R.~G.~Cai,
  Phys.\ Lett.\ B {\bf 634}, 9 (2006)
  [arXiv:astro-ph/0512018];
  J.~Q.~Xia, B.~Feng and X.~M.~Zhang,
  Mod.\ Phys.\ Lett.\ A {\bf 20}, 2409 (2005)
  [arXiv:astro-ph/0411501];
    Z.~K.~Guo, Y.~S.~Piao, X.~M.~Zhang and Y.~Z.~Zhang,
  Phys.\ Lett.\ B {\bf 608}, 177 (2005)
  [arXiv:astro-ph/0410654];
  B.~Feng, M.~Li, Y.~S.~Piao and X.~Zhang,
  Phys.\ Lett.\ B {\bf 634}, 101 (2006)
  [arXiv:astro-ph/0407432];
  Y. Cai, H. Li, Y. Piao, and X. Zhang, gr-qc/0609039;
   R.~Lazkoz and G.~Leon,
  Phys.\ Lett.\ B {\bf 638}, 303 (2006)
  [arXiv:astro-ph/0602590];
  W.~Zhao,
  Phys.\ Rev.\ D {\bf 73} (2006) 123509
  [arXiv:astro-ph/0604460].

  \bibitem{III}
   J.~Q.~Xia, G.~B.~Zhao, B.~Feng, H.~Li and X.~Zhang,
  Phys.\ Rev.\ D {\bf 73}, 063521 (2006)
  [arXiv:astro-ph/0511625];
  G.~B.~Zhao, J.~Q.~Xia, B.~Feng and X.~Zhang,
    arXiv:astro-ph/0603621.

  \bibitem{interact}
  H.~S.~Zhang and Z.~H.~Zhu,
  Phys.\ Rev.\ D {\bf 73}, 043518 (2006);
 H.~Wei and R.~G.~Cai,
  Phys.\ Rev.\ D {\bf 73}, 083002 (2006)


 \bibitem{Li}
  M.~z.~Li, B.~Feng and X.~m.~Zhang,
  arXiv:hep-ph/0503268.
\bibitem{Binflation}
 A.~Anisimov, E.~Babichev and A.~Vikman,
  arXiv:astro-ph/0504560.



   \bibitem{hs1}
 R.~G.~Cai, H.~s.~Zhang and A.~Wang,
  Commun.\ Theor.\ Phys.\  {\bf 44}, 948 (2005)
  [arXiv:hep-th/0505186].
\bibitem{kamen}
  A.A. Andrianov, F. Cannata and A. Y. Kamenshchik,
   Phys.Rev.D72, 043531,2005;
   F.Cannata and A. Y. Kamenshchik,
  , gr-qc/0603129; Pantelis S. Apostolopoulos and Nikolaos Tetradis
 , Phys. Rev. D 74 (2006) 064021.

 \bibitem{maarten}
 R. Maartens,
 gr-qc/0312059.

\bibitem{sahni}
    V. Sahni and Y. Shtanov,  JCAP 0311 (2003) 014, astro-ph/0202346.

 \bibitem{dgpmodel}
 G. Dvali, G. Gabadadze, M. Porrati, Phys. Lett. B485 (2000) 208
 ,hep-th/0005016; G. Dvali and G. Gabadadze, \prd
 {\bf 63},
 065007 (2001); A. Lue, astro-ph/0510068.

\bibitem{dgpcosmology}
C. Deffayet, Phys. Lett. B 502, 199 (2001); C. Deffayet, G.Dvali,
and G. Gabadadze, Phys. Rev. D 65, 044023 (2002); C.Deffayet, S. J.
Landau, J. Raux, M. Zaldarriaga, and P. Astier, Phys. Rev. D 66,
024019 (2002);
 \bibitem{ruth}
 L.~P.~Chimento, R.~Lazkoz, R.~Maartens and I.~Quiros,
  JCAP {\bf 0609}, 004 (2006)
  [arXiv:astro-ph/0605450];
  R.~Lazkoz, R.~Maartens and E.~Majerotto,
  Phys.\ Rev.\ D {\bf 74}, 083510 (2006)
  [arXiv:astro-ph/0605701].
   \bibitem{inho}
  S.~Nojiri and S.~D.~Odintsov,
  Phys.\ Rev.\ D {\bf 72}, 023003 (2005)
  [arXiv:hep-th/0505215].
  \bibitem{frwg}
  S.~Nojiri and S.~D.~Odintsov,
  Phys.\ Rev.\ D {\bf 70}, 103522 (2004)
  [arXiv:hep-th/0408170].


 \bibitem{origin}
 D.M. Capper, Nuovo Cim. A 25 1975 29.

  \bibitem{hs2}
  R.~G.~Cai and H.~s.~Zhang,
  JCAP {\bf 0408}, 017 (2004)
  [arXiv:hep-th/0403234];
  \bibitem{pz}
  E.~Papantonopoulos and V.~Zamarias,
  JCAP {\bf 0410}, 001 (2004)
  [arXiv:gr-qc/0403090].
   \bibitem{inflaton}
  S.~Capozziello, S.~Nojiri and S.~D.~Odintsov,
  Phys.\ Lett.\ B {\bf 632} (2006) 597
  [arXiv:hep-th/0507182];
  S. Nojiri and S.D. Odintsov, Gen.Rel.Grav. 38 (2006) 1285.

\bibitem{WMAP}
  D.~N.~Spergel {\it et al.},
  arXiv:astro-ph/0603449.

\bibitem{tracker}
 I.~Zlatev, L.~M.~Wang and P.~J.~Steinhardt,
  Phys.\ Rev.\ Lett.\  {\bf 82}, 896 (1999).











\end{thebibliography}
\end{document}